\documentclass[9pt,twocolumn,twoside]{opticajnl}

\journal{ao} % Choose journal (ao,jocn,josaa,josab,ol,optica,pr)
\setboolean{shortarticle}{false}
\usepackage{lineno}
\usepackage{soul}
\title{Quantum computing teaching with CoSpaces}

\author[1]{Francesco Sisini}
\author[2]{Igor Ciminelli}
\author[3]{Fabio Antonio Bovino}

\affil[1]{Scientific Department , Tekamed srl, Via Bellaria 6, 44121 Ferrara}
\affil[2]{School of Disruption | SIDI}
\affil[3]{Quantum Optics Lab, Dept. SBAI, SAPIENZA University of Rome, via Antonio Scarpa 14/16, 00161 Roma }

\affil[*]{Corresponding author: francescomichelesisini@gmail.com}

\begin{abstract}
The first prototypes of quantum computers sparked interest in quantum computing and the basic principles of quantum mechanics. The education project on the physical bases of quantum computing is part of this context, based on the experimental description with virtual methods of the physical implementation of Di Vincenzo's first 5 principles. The computation process is implemented as transformations of qubits encoded in the polarization of optical photons. These transformations are implemented as quantum gates made as 3D virtual objects using Blender. In detail, the models of: Laser Ar +, PBS, HWP / QWP, BBO, APD, SMF, Control electronics are made. With the 3D models, a virtual laboratory has been created within CoSpaces where it is possible to become familiar with the basic processes of quantum computing: production of announced photons, transformation of a qubit, measurement of a qubit, production of entangled photons, transformation of two qubits, measure of two qubits. The realization of physical models to be used in the metaverse could fill the didactic void due to the absence of quantum optics laboratories.
\end{abstract}

\begin{document}

\maketitle

\section{Introduction}

Quantum computing is enjoying great success in different environments, from school to professional and business, thanks in particular to the promise of creating systems that are currently unworkable through traditional information technology, such as secure \cite{bovino} cryptography.
The enthusiasm for this discipline, which in fact is also generating significant market hype, seems to be more linked to marketing phenomena than to its real scientific content and its real applications.
To have a real positive impact, innovation needs widespread diffusion and a growing level of adoption to express its full \cite{Everett} potential. In the case of quantum computing, this assumption today collides with two main obstacles. The first is related to the accessibility to the tools (hardware) that would allow the practical learning of the technology. The second is the complexity of the subject itself, which often discourages the non-specialist public.
This work aims to overcome these obstacles in order to democratize access to the understanding and use of quantum computing, in particular by attracting young minds, in order to increase its adoption and prevent the repetition of what is already happened for traditional information technology, of which the mass public still ignores the basic principles, despite the spread of technological products, while also the specialist public often misinterpreted both the limits and the actual possibilities.
Our studio changes the didactic approach, using CoSpaces as an engagement and learning tool. By introducing quantum computing as one of the different declinations of experimental computation \cite{Horsman}, \cite{Beggs}, instead of as a reality in its own right, the learning path is reversed starting from the experiment and going back to the theory. In practice, instead of presenting quantum computing starting from the quantum theory of computation, we introduce it through some quantum physical systems, made in 3d in a virtual environment, sufficient to implement Di Vincenzo's first 5 criteria. Systems whose dynamic evolution produces and processes qubits.

We believe that this approach not only simplifies learning, but is highly attractive to students thanks to the use of the Metaverse and a \textit{gaming} component. For example, in the context of secondary education, where computer science can be introduced in a natural way as the result of the dynamic evolution of a physical system, our project can become a tool that engages young minds and encourages study and interest in the subject. The objective of this study was to identify, analyze and formalize quantum physical systems to create a coherent and uniform presentation on two distinct levels: one is the practical one, created through 3D models of physical systems arranged in a virtual laboratory of a platform. of Metaverse, the other is the descriptive and formal one that uses the natural language of physics. The first floor is mainly aimed at students and enthusiasts who find there a concretization of ideas and concepts that are often disclosed in approximate terms and that can confuse ideas rather than clarify them. The second is aimed at teachers, disseminators and trainers, who through the formal description can spend themselves in clear and constructive dissertations based on theoretical foundations.

%%Sisini

\subsection{DiVincenzo's criteria}
\label{sec:divincenzo}
The purpose of a quantum computer is to physically implement the states of the qubits and the unitary transformations that operate on them where the term "physically" means that there must really exist, not only in mathematical formulas, an instrument that has qubits and of devices to modify them. Quantum computing is therefore based on the design and construction of a real \textit{physical} system composed of qubits registers and instrumentation acting on them.\\
As can be expected this is a very complex conceptual and technological challenge.
To address this challenge by transforming it into the solution of a problem with an analytical method, the physicist DiVicenzo proposed a list of requirements for building a physical system capable of quantum processing, that is, a quantum computer.
The criteria are divided into two groups, the first concern the computation:
\begin {itemize}
   \item A system of well characterized and scalable qubits in size
   \item The ability to initialize (set) the state of the qubits
   \item Long decoherence times (loss of coherence)
   \item A universal quantum gates system
   \item The ability to measure each qubit
\end{itemize}
The second relate to communication
\begin {itemize}
     \item Possibility of transforming calculation qubits into information exchange qubits
     \item Ability to reliably transmit qubits
\end{itemize}
In this paper, in accordance with the stated objectives, we will identify a set of physical systems that implement DiVincezo's first five principles.

\section{Materials and Methods}
\label{sec:mem}
The technological solutions implemented so far for the realization of qubits range from optics to the physics of matter.
For the purposes of this work, we have chosen to implement qubits as the polarization state of a single photon and therefore to implement the physical production and processing systems as optical systems. Five different systems have been described in detail, listed below:
\begin {enumerate}
    \item Production of an announced photon (heralded)
    \item Transformation of the polarization state of a single photon
    \item Projective measure of the polarization state of a single photon
    \item Production of two photons in an entangled state
    \item Realization of a transformation of the state of two photons (C-Not)
\end {enumerate}
Each of the proposed systems has been extrapolated from the description of an experimental setup published in peer reviewed journals and has been adapted by replacing the instrumentation actually used with an idealized instrumentation. Overall the tools used for all the experiments are:
\begin{itemize}
    \item Laser Ar+
    \item Polarizing beam splitter (PBS)
    \item Prisma ottico
    \item Quarter wave plate (QWP)
    \item Half wave plate (HWP)
    \item Avalange photo diode (APD)
    \item Single photon counter
    \item Borate crystal of Barium: BaB$_2$O$_4$ (BBO)
    \item Single mode fiber (SMF)
\end{itemize}
Each of the elements listed above was built as a 3D model using Blender software (\url{https://www.blender.org/}). The individual elements were then arranged in different setups to create the 5 physical systems shown in the list above.
For each of the five physical systems described, a formal description was created that describes the main physical phenomena involved, a 3D scene with the experimental setup and a \emph{space} on CoSpaces (\url{https://edu.cospaces.io/Universe}): a platform for teaching and developing augmented reality (AR) and virtual reality (VR). The CoSapces environment allows you to develop interactive 3D projects using 3D objects and application code. Projects can result in non-interactive 3D videos or in interactive applications such as simulations or video games.

\section{Results}
\subsection{Heralded photon production}
\label{sec:spdc}
\subsubsection{Physical principles}
\begin{figure}[htbp]
\centering
\fbox{\includegraphics[width=\linewidth]{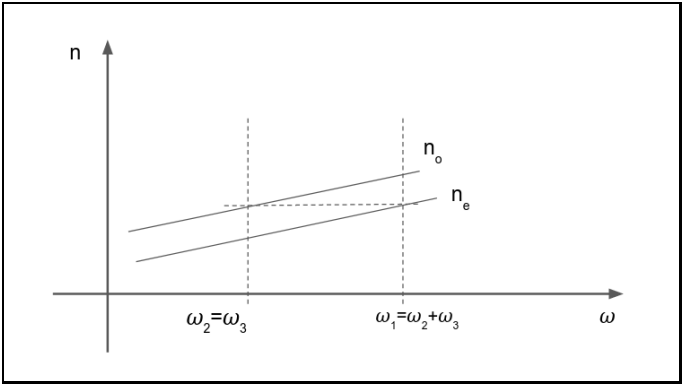}}
\caption{The figure shows the graph of the ordinary and extraordinary refractive indices (n$_o $ and n$_e $). The existence of two different refractive indices allows the conservation of the moment in the SPDC TYPE-I process.}
\label{fig:momentum}
\end{figure}

An Argon ion laser is used to produce a photon beam with wavelength $ \lambda = 351 $ nm. % The beam produced is purged of fluorescence and / or other wavelengths of Ar $ ^ + $ using a refractor prism.
The laser beam then passes through a PBS to obtain a vertically polarized transmitted beam and a horizontally polarized reflected beam at the output.
The polarization angle of the laser beam is further corrected by a HWP, this to allow greater accuracy than what can be obtained by directly rotating the PBS. The laser then reaches a BBO crystal which has ordinary and extraordinary refractive indices of 1.6776 and 1.5534 respectively. The birefringence of the BBO allows the absorption of one photon of the laser ($ k_1, \omega_1) $ and the relative emission of two photons ($ k_2, \omega_2) $ and ($ k_3, \omega_3) $ by means of a Spontaneus Parametric Down Conversion (SPDC) type I. The polarization state of the two emitted photons must be the same and complementary to the state of the absorbed photon, this for conservation of energy and moment, in fact we have that the conservation of the moment:
\begin{equation}
    k_2+k_3=k_1
\end{equation}
then
\begin{equation}
    n(\omega_3)\frac{\omega_3}{c}+n(\omega_2)\frac{\omega_2}{c}=n(\omega_1)\frac{\omega_1}{c}
    \label{eq:momentum}
\end{equation}
The equation \ref{eq:momentum} can only be satisfied if the two emitted photons have the same polarization, as seen in figure \ref{fig:momentum}.
The two photons emitted following the SPDC therefore have the same polarization and are emitted at a known angle. One of the two photons is then led to a detector (APD) while the other is led to other stages of transformation that will be seen below. The detection of the first photon (herald photon) therefore announces the presence of a photon of known polarization that can be used for further processing. \\
This system realizes the first two criteria of DiVincenzo establishing in fact the use of the polarization of the photon as a qubit and the possibility of initializing the state of the qubit. \\
\subsubsection{Implementation in Blender}
\begin{figure}[htbp]
\centering
\fbox{\includegraphics[width=\linewidth]{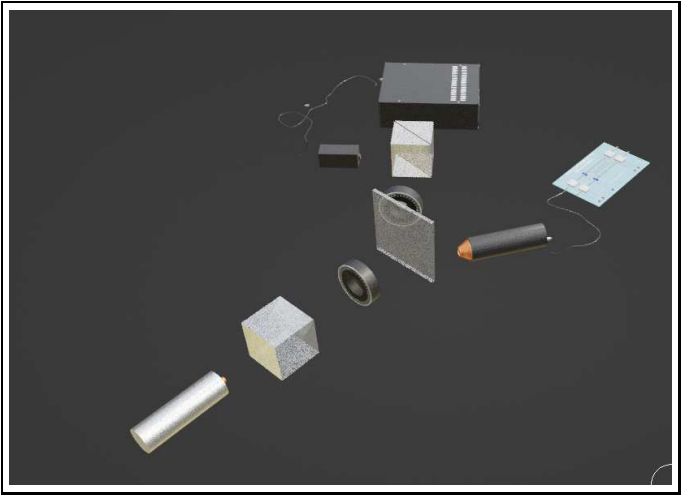}}
\caption{The figure shows the realization of the experimental setup in Blender.}
\label{fig:s1}
\end{figure}

The complete diagram of the physical system is shown in figure \ref{fig:s1}. The experimental setup was basically taken from \cite{Kwiat}.

\subsubsection{Virtual reality application}

\begin{figure}[htbp]
\centering
\fbox{\includegraphics[width=\linewidth]{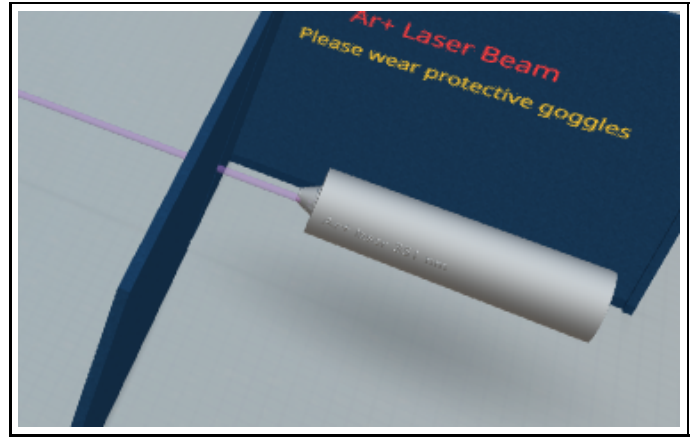}}
\caption{The figure shows the simulation of the laser beam implemented as a purple cylinder exiting the laser generator.}
\label{fig:laser}
\end{figure}

\begin{figure}[htbp]
\centering
\fbox{\includegraphics[width=\linewidth]{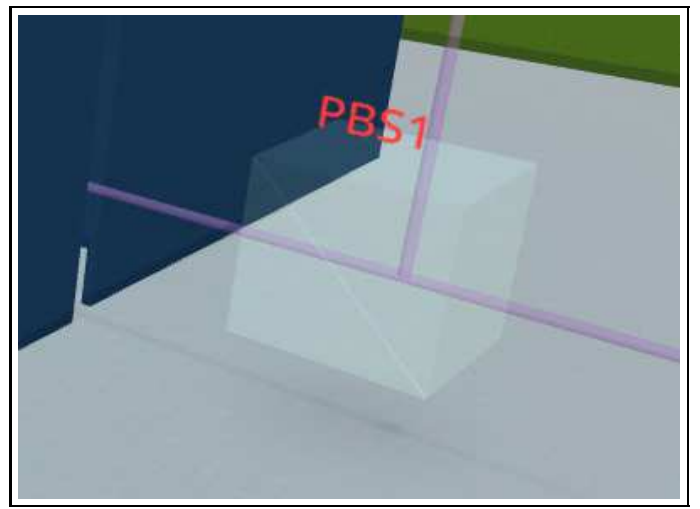}}
\caption{The figure shows the separation of the laser into two polarization beams perpendicular to each other.}
\label{fig:PBS}
\end{figure}

The objects developed with Blender were exported in GL Transmission Format Binary file (.glb) and imported on CoSpaces. The objects were imported one by one separately so that it was possible to separately manage the events generated by the interaction with the user (mouse click, etc.). A laser beam and two photons were also modeled, the first as a cylinder and the second as two spheres. The laser beam leaves the Ar$_+$ laser generator (see figure \ref{fig:laser}) and separates into two components at the first PBS (see figure \ref{fig:PBS}). The two spheres represented the photons are generated at the BBO and then reach the one APD and the other SMF. \\
Using the objects described above, a simulation of the production process of the announced photons was implemented. The simulation is descriptive in fact the angle between the two outgoing photons and their production rate are not respected, but on the whole the main processes involved in the experiment are correctly represented. \\
The simulation at the point above was used to develop an interactive path in which the user must correctly configure the experiment to observe the production of heralded (announced) photons and their correct detection. In detail, the user must first activate the laser beam, then adjust the two PBS to the correct angle.
\\
The objects were physically arranged according to the scheme in figure \ref{fig:s1}. Together with the objects of the physical experiment, a floor and walls were arranged so that the experiment is placed in a physical environment in which the various components are arranged in separate rooms, to create a scenic effect and a playful setting. Articles of interest have been hung on the walls and can be viewed and clicked to access the original documents. Using the partition walls, the names of the physical devices and the explanation of the role in the experiment were entered. \\
The space on CoSpaces of this experiment has been named "Heralded photon production" and is publicly reachable from the address \url{https://edu.cospaces.io/ZQA-LGN}.

The experiment requires activating the laser by clicking on it and positioning the HWP at 0°, then clicking on the BBO to view the emission of the two photons due to SPDC TYPE I.

\begin{figure}[htbp]
\centering
\fbox{\includegraphics[width=\linewidth]{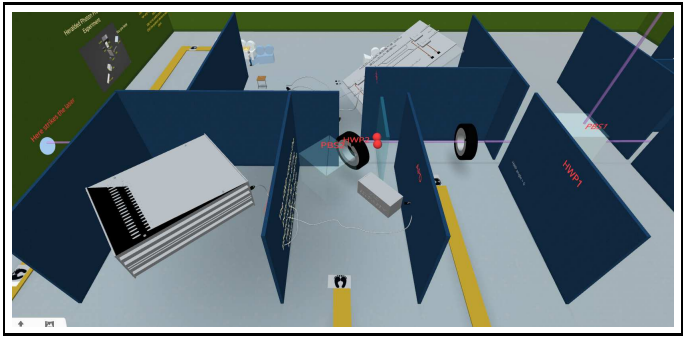}}
\caption{The figure shows the implementation of the experimental setup in CoSpaces.}
\label{fig:cospaces_1}
\end{figure}

\subsection{Transformation of the polarization state of a single photon}
\subsubsection{Physical principles}
\begin{figure}[htbp]
\centering
\fbox{\includegraphics[width=\linewidth]{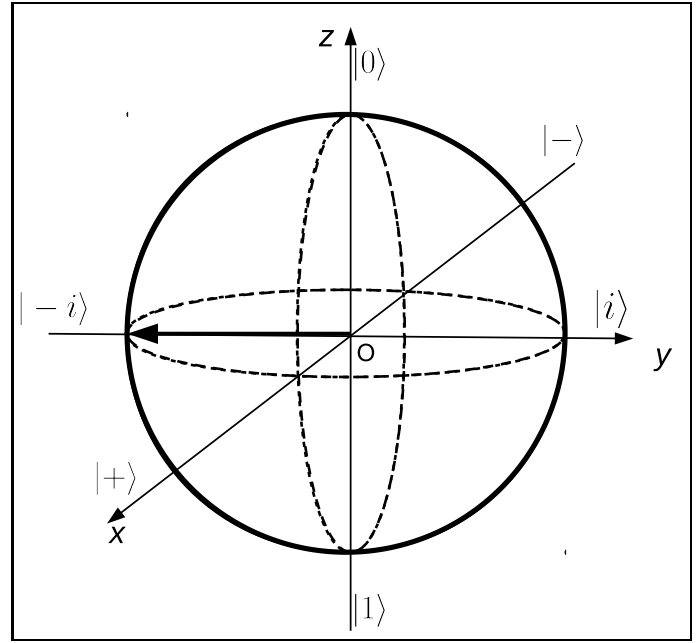}}
\caption{The figure shows a qubit in the state $ | -i \rangle $ in the Bloch sphere.}
\label{fig:bloch}
\end{figure}

The polarization state $ |\psi  \rangle $ of a photon can be expressed as
\begin{equation}
  |\psi\rangle = \alpha |0\rangle+\beta |1\rangle
  \label{eq:stereo0}
\end{equation} 
where the coefficients $ \alpha $ and $ \beta $ are complex numbers. For a pure state, the condition $ | \alpha | ^ 2 + | \beta | ^ 2 = 1 $ allows to represent the polarization state (qubit) on a spherical surface plus a point, this representation is called Bloch sphere ( see figure \ref{fig:bloch}). All possible transformations of a qubit can be seen as rotations of the unit vector representing the point on the Bloch sphere. Since an arbitrary rotation in $ R^3 $ can be decomposed into three independent rotations around two non-parallel axes, it is possible to realize any rotation of the polarization by three successive rotations around two axes. To carry out this transformation, waveplats were therefore used that perform rotations around two axes of the Bloch sphere.
If you choose the $ x $ axis and the $ y $ axis as axes, then you can use a QWP for the rotations around the $ y $ axis and a HWP for the rotations around the $ x $ axis of the sphere by Bloch.
  In the system proposed here, two QWPs and one HWP are used to produce an arbitrary transformation of the polarization state of a photon. The three devices are aligned on a single axis according to the QHQ scheme, i.e. a QWP, a HWP and a QWP. Rotating them respectively by $ \alpha $, $ \beta $ and $ \gamma $ we obtain the unitary transformation $ U $ which corresponds to a rotation around the $ n $ axis in the Bloch sphere of the photon polarization:
\begin{equation}
    U_{n}=R_{p}(\alpha)R_{q}(\beta)R_{p}(\gamma)
\end{equation}
which corresponds to a transformation of the state of the qubit.
where $ n $ denotes the axis of rotation, $ R $ is the rotation matrix, and $ p $ and $ q $ are two non-parallel directions. \\
This physical system actually implements part of point four of DiVincenzo's criteria, providing the ability to perform any unitary transformation $ | \psi \rangle \rightarrow | \psi^{'} \rangle $ on a single qubits, thus providing half of the system of universal quantum gates.
\begin{figure}[htbp]
\centering
\fbox{\includegraphics[width=\linewidth]{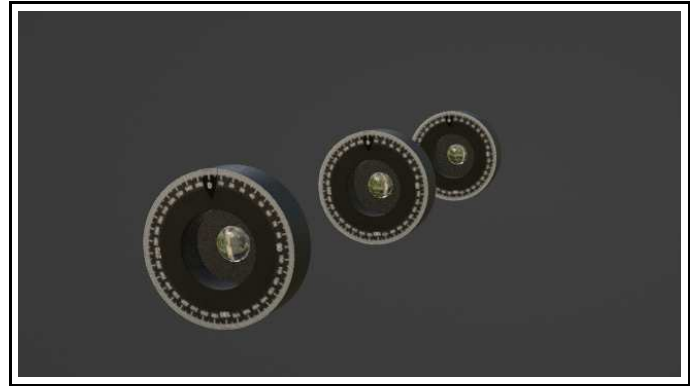}}
\caption{The figure shows three waveplates made with Blender.}
\label{fig:qhq}
\end{figure}
The complete theoretical description and implementation examples in \cite{Peters},\cite{Langford},\cite{Sit}
\subsubsection{Realization in Blender}
The system was built on Blender by aligning three HWPs with each other. From a graphic point of view, in fact, there is no difference between a HWP and a QWP. The modeling result is visible in \ref{fig:qhq}.
\subsubsection{Virtual reality application}
\begin{figure}[htbp]
\centering
\fbox{\includegraphics[width=\linewidth]{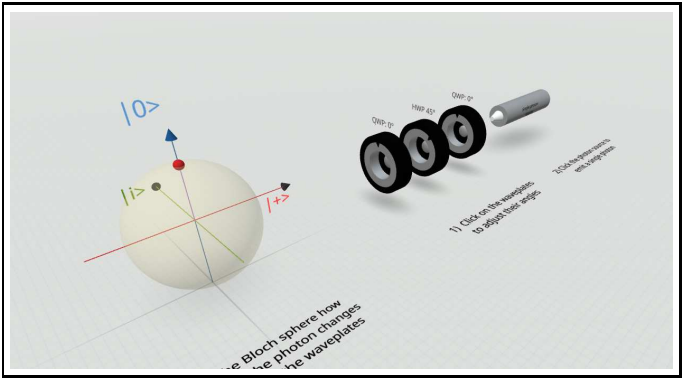}}
\caption{The figure shows the quantum gates operating on CospaCes.}
\label{fig:SingleQubit}
\end{figure}

Wave plates made with Blender have been uploaded to a dedicated space on CospaCes, reachable at \url{https://edu.cospaces.io/KYG-EJM}. To their right an object has been arranged that represents a source of a single initialized photon, in practice it summarizes what we saw in the previous experiment in a single object that integrates all the necessary components. The source is aligned with the optical axis of the three wave plates in this way the photons emitted by it can pass through all three. For this purpose, the code that simulates the emission of a horizontally polarized photon in response to the click on the object has been implemented. The angle of each of the three wave plates can be changed as desired in a 5° step. To the left of the plates there is an object that represents the direction of the qubit (polarization of the photon) in the Bloch sphere. As the photon passes through each plate, the qubit rotates in the Bloch sphere showing the action of the three plates which then create a quantum gate. The system is represented in figure \ref{fig:SingleQubit}.

\subsection{Projective measure of the polarization state of a single photon}
\subsubsection{Physical principles}
The density matrix of a quantum system can be experimentally deduced through a set of projective measures. For a two-level system such as the (photonic) qubit, which is in a pure state, the state of each photon can be written as: $ | \psi \rangle = \alpha | H \rangle + \beta | V \rangle $ and that the square modules of the coefficients $ \alpha $ and $ \beta $ can be determined through two projective measures:
\begin{equation}
|\alpha|^2=|\langle H |\psi \rangle|^2
\end{equation}
e
\begin{equation}
|\beta|^2=|\langle V |\psi \rangle|^2
\end{equation}
If the detectors (APD) used for the two measurements are ideally perfect, the condition $ | \alpha | ^ 2 + | \beta | ^ 2 = 1 $ must be valid, therefore it is sufficient to perform one of the two measurements to also know the 'other. However, the knowledge of the square modules of $ \alpha $ and $ \beta $ is not sufficient to determine the knowledge of the state of the qubit since they are complex numbers $ \alpha = a + ib $ and $ \beta = c + id $.
There are therefore in total four unknowns which must be determined with four equations. As we have seen, one equation is already given, the other three instead are obtained through projective measures of the state of the photon. For example by measuring $ | \psi \rangle $ with respect to the two states $ H $ and $ V $,
then $ |\psi^{'} \rangle $ and $ | \psi^{''} \rangle $ with respect to the same basic states. Obviously, it is necessary to have a set of photons that are in the same state, because for obvious reasons it would not be possible or useful to perform multiple measurements on the same photon. The states $ | \psi^{'} \rangle $ and $ | \psi^{''} \rangle $ can be obtained by rotating $ | \psi \rangle $ using a combination of a QWP and a HWP. In particular it has:
\begin{equation}
    |\psi\rangle =HWP(0)QWP(0)|\psi\rangle 
\end{equation}
\begin{equation}
    |\psi^{'}\rangle =HWP(\frac{\pi}{8})QWP(\frac{\pi}{4})|\psi\rangle 
\end{equation}
e
\begin{equation}
    |\psi^{''}\rangle =HWP(\frac{\pi}{8})QWP(0)|\psi\rangle 
\end{equation}
To realize this measurement system, a QWP and a HWP are arranged on the same optical axis of a PBS. Two APDs are arranged one at the output ports of the PBS.
\cite{Omer}
\begin{figure}[htbp]
\centering
\fbox{\includegraphics[width=\linewidth]{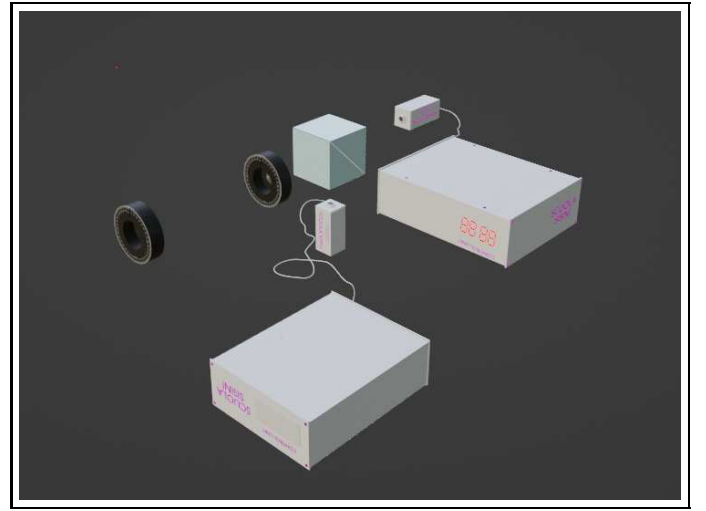}}
\caption{The figure shows two waveplates made with Blender aligned to a PBS. There are two APDs at the ports of the PBS.}
\label{fig:misura_proiettiva}
\end{figure}
\subsubsection{Realization in Blender}
The system was built on Blender using the HWP, PBS, APD and single photon cuounter model. The modeling result is visible in \ref{fig:misura_proiettiva}.

\subsubsection{Virtual reality application}
\begin{figure}[htbp]
\centering
\fbox{\includegraphics[width=\linewidth]{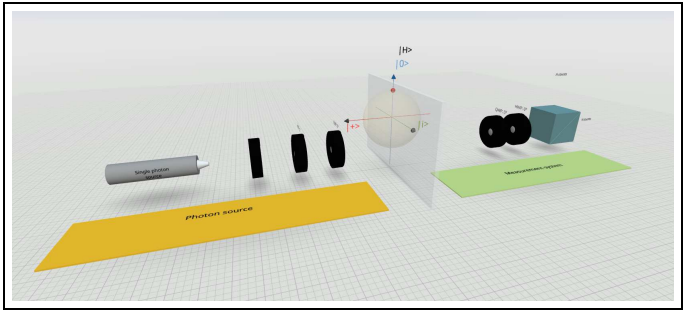}}
\caption{The figure shows a projective measurement system built and operational on CospaCes}
\label{fig:projective_measurement}
\end{figure}

The realization of this experiment on CospaCes (\url {https://edu.cospaces.io/YCB-QWY}) allows the user to configure the two wave plates described in the previous paragraph by changing the angle of the optical axis as desired . To the left of the measurement system is an idealized single photon source that produces only one photon at a time in the $ | H \rangle $ polarization state. On the same optical axis there are three waveplates with the QHQ scheme that allow you to change the state of the qubit at will. Clicking on the photon source a photon is emitted. The system then calculates the polarization state of the photon on the basis of the five wave plates present on its trajectory and shows on the Bloch sphere the state of the qubit in correspondence with the plate crossed. At the end of the optical path there is a PBS. When the photon reaches the PBS the system calculates the probability that the photon will be detected at the horizontal or vertical gate. The system also takes into account the number of photons detected along both gates. The system is shown in the figure \ref{fig:projective_measurement}.

\subsection{Production of two photons in an entangled state}
\subsubsection{Physical principles}
In the section \ref{sec:spdc} we illustrated the SPDC TYPE I process showing how a photon coming from the laser pump and polarized $ | V \rangle $ can produce two polarized photons in the state $ | H, H \rangle $. If the photon is in a combination $ | \psi \rangle = \alpha | H \rangle + \beta | V \rangle $ the process will still occur but with a probability scaled by the factor $ | \beta | ^ 2 $. To produce a pair of entangled photons the idea is to align two BBO crystals. The identically cut crystals are oriented with their optical axes
 aligned on perpendicular planes, i.e. the first optical axis of the crystal and the radius of the pump define the vertical plane while the second optical axis and the pump define the horizontal one. The polarization of the laser pump is set to $ | \psi \rangle = \frac {1} {\sqrt {2}} (| H \rangle + | V \rangle) $ by aligning a HWP between the crystals and a PBS. The result is the production of two photons in an entangled state $ \frac {1} {\sqrt {2}} (| H, H \rangle + | V, V \rangle) $. The experimental setup is described in \cite{Kwiat}. This point does not specifically implement one of DiVincenzo's criteria but is necessary for the realization of the C-Non quantum gate.

\subsubsection{Realization in Blender}
\begin{figure}[htbp]
\centering
\fbox{\includegraphics[width=\linewidth]{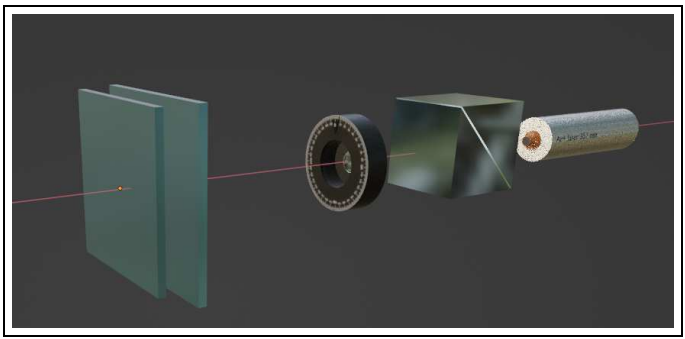}}
\caption{The figure shows the configuration of two BBO crystals for the production of entangled photons.}
\label{fig:entangled}
\end{figure}

The complete diagram of the physical system is shown in figure \ref{fig:entangled}. The experimental setup was basically taken from\cite{Kwiat}.
\subsubsection{Virtual reality application}
\begin{figure}[htbp]
\centering
\fbox{\includegraphics[width=\linewidth]{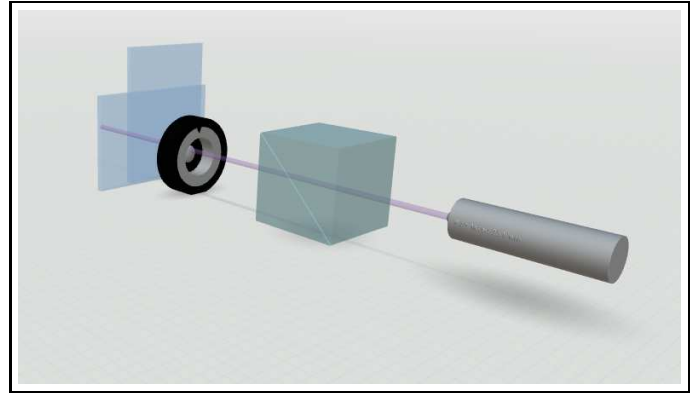}}
\caption{The figure shows two BBO crystals arranged one with the optical axis perpendicular to the other on CoSpaces}
\label{fig:enta}
\end{figure}

The implementation of this experiment on CoSpaces allows the user to produce two photons in the state $ | \Psi^{-} \rangle $. The system is very simple. The polarization of the laser pump can first be selected by means of the PBS placed between it and the pair of BBO crystals. Between the PBS and them is then placed a HWP for the fine correction of the polarization angle which to maximize the production of a pair of photons for SPDC must be perpendicular to the optical axis of one of the two crystals and parallel to the other. The system is shown in the figure \ref{fig:enta}

\subsection{Realization of a transformation of the state of two photons (C-Not)}
\subsubsection{Physical principles}
To satisfy DiVincenzo's fourth criterion it is necessary to implement at least one quantum gate acting on two qubits. C-Not is the standard gate in most quantum computer implementations based on quantm gates. The C-Not foresees that the exit state of the two qubits entering the gate depends on their entry state which means an interaction between the two qubits or an interaction between the qubits and the environment. The physical implementation of C-Not with photonic qubits poses the problem of non-interaction between photons and the difficulty of obtaining non-linear effects using single photon states. The C-Not can however be implemented with linear optical transformations (KLM) \cite {KLM} where the non-linearity term is instead introduced by a measurement operation. A very efficient way to implement C-Not with linear optics consists in exploiting two auxiliary (ancillary) photons produced in the entangled state. This type of C-Not is said to be heralded or heralded because the correct transformation of the \emph {target} qubit is probabilistic but is announced by the detection of ancillary photons. A complete implementation of this circuit is shown in \cite {Zeuner}, \cite {Pittman}.
The system consists of two PBS (PBS $ _1 $ and PBS $ _2 $), one in the H / V base and the other in the D/A base (i.e. given by the states $ | + \rangle $ and $ | - \rangle $ ). The two qubits controller (c) and target (t) in input $ c_ {in} = \alpha | H \rangle + \beta | V \rangle $ and $ t_ {in} = \gamma | H \rangle + \delta | V \rangle $ each enter a PBS through one port, while one of two ancillary photons that are in an entangled state enters the other. Two projective measurement systems are arranged in correspondence to two of the four total output ports of the system: the first of a PBS in the D/A base and a pair of APDs and the second identical but in the H/V base. \\

An explanation, even a superficial one, of the experiment is necessary for the didactic purposes pursued by this work. We then divide the experiment into two parts. In the first we consider the two ancillary photons defined by a factorizable (non-entangled) state. In this scenario,
each of the two beam splitters (PBS $ _1 $ and PBS $ _2 $) works as a parity detector, i.e. the status transmitted along the output port ($ c_ {out} $ for PBS $ _1 $ and $ t_ {out} $ for PBS $ _2 $) is enabled only if the relevant detector receives 1AO1 (one and only one) photons. In this case (1AO1), if we consider the two PBSs separately, it is easy to show \cite {Pittman} that the exit state of each PBS corresponds to the entry state, so $ c_ {out} = c_ {in} $ and $ t_ {out} = t_ {in} $. Obviously, in the process, trace of the incoming photons is lost as they interfere with the two ancillary photons, what is really transferred is the state of polarization (the qubit) and not the photon itself which instead loses its distinction. \footnote {We insist on this point because for application reasons the indistinguishability of particles has not been transferred to the VR application. For this reason it is important to have clear the difference between what is shown in VR and what is correct in terms of quantum theory.} Proceeding in the second part we consider the complete experiment and therefore of two ancillary photons produced in an entangled state, for example the state $ | \Psi - \rangle $. In a completely analogous way, it is shown that the state transferred to the output ports $ | c_ {out}, c_ {out} \rangle = \alpha | H \rangle (\gamma | H \rangle + \delta | V \rangle) + \beta | V \rangle (\gamma | V \rangle + \delta | H \rangle) $, i.e. exactly the theoretical action of a C-Not.

\subsubsection{Realization in Blender}
\begin{figure}[htbp]
\centering
\fbox{\includegraphics[width=\linewidth]{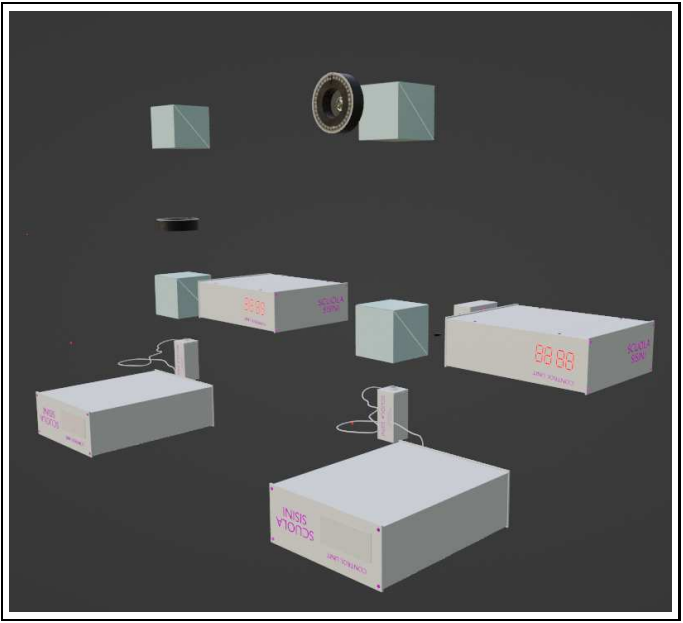}}
\caption{The figure shows the realization of a C-not with linear optics components.}
\label{fig:CNOT}
\end{figure}

The complete diagram of the physical system is shown in figure \ref{fig:CNOT}. The experimental setup was basically taken from \cite{Zeuner}.
\subsubsection{Virtual reality application}
\begin{figure}[htbp]
\centering
\fbox{\includegraphics[width=\linewidth]{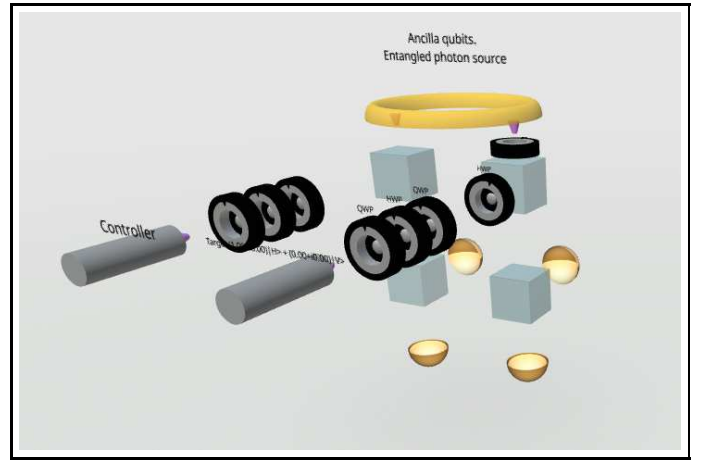}}
\caption{The figure shows the realization of a C-not with linear optics components on CoSpace.}
\label{fig:C_NOT}
\end{figure}

The system was built on CospaCes similar to previous systems (\url {https://edu.cospaces.io/STC-NGC}). There are two single photon generators representing the control qubit $ c_ {in} $ and the target qubit $ t_ {in} $. On the flight line of both are arranged three waveplates in QHQ configuration that allow you to transform, by clicking on the wave plates, the two qubis at will before crossing the relative PBS.
The two PBS based D / A are made by placing a 22.5 ° HWP on their entrance doors. The entangled photon source is represented with a ring generating two photons in the state $ | \Psi - \rangle $ in the gates $ a ^ 1_ {in} $ and $ a ^ 2_ {in} $.
In correspondence to the two output ports $ a ^ 1_ {out} $ and $ a ^ 2_ {out} $ there are two other PBS, one based on D / A and the other based on H / V, each of which has in correspondence to the output ports two detectors $ D_1 $, $ D_2 $ for the first and $ D_3 $ and $ D_4 $. \\
 In total there are therefore 4 photons. Since a true indistinguishability of the spheres used to represent photons is not possible, it was decided to proceed with a semi-classical representation of the experiment in which the photons are distinguishable and distinguished by the color of the relative sphere that represents them. In this way the state of the polarization of the transmitted photon is $ | H \rangle $ or $ | V \rangle $, so in general the input state does not coincide, which can be a superposition of the base states. The behavior of C-Not is however respected if we consider not to use the output status for a subsequent computation but to perform a projective measurement of the qubits $ c_ {out} $ and $ t_ {out} $ in output. \\
 To start the simulation, click on the source labeled "Target" and activate the system that simultaneously emits the four photons $ c_ {in} $, $ t_ {in} $, $ a ^ 1_ {out} $ and $ a ^ 2_ {out} $, as specified, each of a different color. Their state is calculated in a deterministic way with regard to unit transformations (wave plates and beam splitter) and with a pseudo-random function for projective measurements on the detectors. When the condition of 1AO1 is fulfilled, that is when exactly one count is recorded on the pair $ D_1 / D_2 $ and one on the pair $ D_3 / D_4 $ then the system \textit {announces} the proper functioning of the door, allowing the use of the two qubits $ c_ {out} $ and $ t_ {out} $. The system is shown in the figure \ref{fig:C_NOT}.

\section{Discussion}
In our simulation we introduced photons as visible objects, that is, as if they had their own shape and dimension. It is clear that giving shape and size to elementary particles is always a gamble, but it is even more so in the case of the photon since it has no mass and propagates at the speed of light. Beyond that we do not forget Dirac's statement that he writes that the main objective of physics is not to
provide a picture of nature, but provide the laws that
they govern the known phenomena that can lead to the discovery of
new phenomena. \\
Note that Dirac writes exactly:

\ldots{}the main object of physical science is not the provision of
picture,\ldots{}

highlighting the \textbf {no need} to picture the world
quantum.\\
Dirac had a good reason to support this thought since yes
it was assumed the responsibility to overturn the view that physics
had had until the early 1920s. On the other hand, our goal is on the one hand to show a 3D representation of quantum computing experiments that can be inserted into the Metaverso platform to improve the teaching of this topic, on the other hand it is to provide a pictorial representation of what it also happens behind the scenes to those without skills in quantum mechanics. For this reason we think it is acceptable to depart from Dirac's \emph{no picture principle} in well-defined cases. In any case, we believe that the discussion must remain open and that the topic is far from being considered concluded. \\
A second controversial aspect concerns the semi-classic method with which we simulated the C-Not door. This aspect is particularly interesting on the didactic level because it leads the discussion to one of the first works in reference to the quntistic computation by the physicist R. Feynman \cite{Feynman}, i.e. the doubt whether it is possible to simulate or emulate a quantum process on a \textit{classic} computer. The delicate but interesting topic is dealt with in \cite{Sisini} soon published on arxiv.

\section{Conclusions}
\begin{figure}[htbp]
\centering
\fbox{\includegraphics[width=\linewidth]{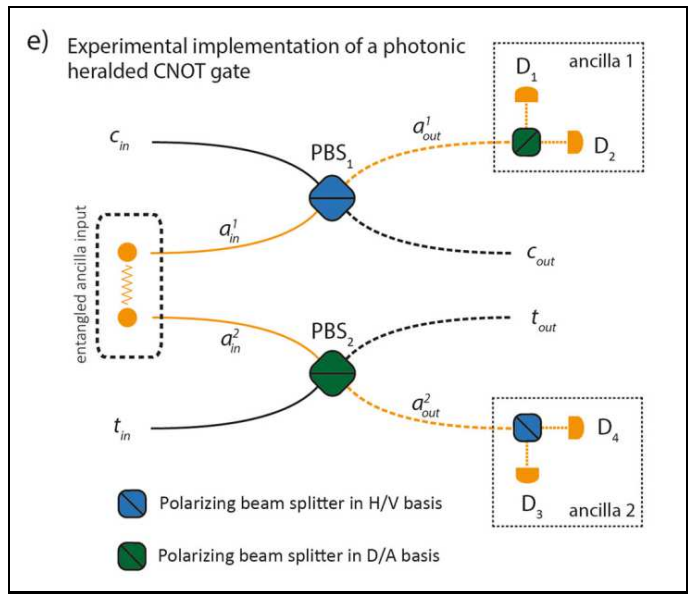}}
\caption{The figure shows the diagram of a c-not heralded. The image is from \cite{Zeuner}.}
\label{fig:schema}
\end{figure}

The goal of identifying, analyzing and formalizing five linear optics experiments sufficient to introduce the basics of quantum computing has been achieved. The experiments were explained and analyzed on a formal level, but above all they were implemented in terms of objects in VR. The result of the effort made in this direction is evident if we compare the content of the original articles with what we modeled. For example in figure \ref{fig:schema} the original diagram of the C-Not is shown as presented in\cite{Zeuner}.
\smallskip
For those who have experience in an optics laboratory, this scheme is obviously evocative of the underlying physical system and therefore is sufficiently explanatory of the contents it must convey. The same scheme, however, is hermetic for a non-expert reader and therefore of little didactic utility if not further interpreted. Our contribution was precisely in transforming schemes like this into 3D models represented in VR that are in direct relationship with physical objects that concretely carry out the physical transformations of quantum computing. Furthermore, the development of animations with which the user / student can interact allows to actively observe the dynamic development of systems over time and not only to imagine it based on the system equations. The last fundamental aspect of this work is to have developed these systems so that they can soon be inserted into a Metaverse platform as CospaCes itself is about to be or like other platforms already on the market. The placement of these experiments in the Metaverse will allow to interact in multy-player modality with the experiments themselves. In practice, students and researchers / teachers will be able to interact simultaneously on the same experiment allowing both didactic and discussion activities.

\end{document}